\documentstyle[12pt]{article}
\begin{document}
\begin{center}
{\bf VARIATIONAL CALCULATIONS OF THE $^{12}C$ NUCLEUS STRUCTURE IN A
3$\alpha$ MODEL USING A DEEP POTENTIAL WITH FORBIDDEN STATES }\\
\vspace*{0.5cm} E.M. Tursunov\\
\end{center}
 \begin{center}
  Institute of Nuclear Physics,702132, Ulugbek, Tashkent, Uzbekistan
 Tel.: +998-712-60-60-01; Fax: +998-712-64-25-90;
 E-mail:tursune@inp.uz
\end{center}
\begin{abstract}
   \par The energy spectrum of the $^{12}C$ nucleus with
  $(J^{\pi}, T)=(0^+,0)$ and $(2^+,0)$
is investigated in the framework of the multicluster dynamical model
by using a deep $\alpha \alpha$-potential with forbidden states in the S
and D waves. A very high sensitivity of the compact ground and first excited $2^+_1$
states energy levels to the description of the two-body forbidden states
wave functions has been estabilished. It is shown also that the chosen
method of orthogonalizing pseudopotentials yields convergent results
for the energies of the excited $(0^+_2,0)$ and $(0^+_3,0)$ states of the
$^{12}C$ nucleus with a well developed cluster like structure.
 \end{abstract}
\vskip 5mm
\par {\bf PACS number(s): 21.45.+v; 21.60.Gx; 03.65.Ge}
\vskip 5mm
{\small {\bf Key words: cluster model, Pauli principle, pseudopotential}}
\newpage
\section { Introduction }
 \par
A systematic study of the $^{12}C$ nucleus structure has been successfully
performed on the basis of the microscopic three-alpha models
\cite{hut70,smir74,hor75,tosh80,smid82}. These models reproduce the general
features of the low-lying
$^{12}C$ spectrum.
In the calculations  using the resounating group method (RGM) and
 the orthogonality condition model (OCM) the ground and first excited $2_1^+$
 states energies of the $ ^{12}C$ nucleus are overbound by 7-10 MeV, although
 the energy positions of the $0_2^+, 2_2^+, 3_1^-$ and $1^-$ resonance states
 are reproduced fairly well. In the recent work \cite{cso97}, which used the
 RGM method, the low-lying energy spectrum of this nucleus was described with
 a good accuracy. This work has shown also that the astrophysical significant
 $0_2^+$ state of the carbon is a genuine three-alpha resonance in the
 continuum. This fact is important, since almost all the carbon is produced
 via cupture reaction $ \alpha + ^8Be  \rightarrow ^{12}C+\gamma $, which goes
 through $0_2^+$ resonance of the $^{12}C$ nucleus.
 \par Although the microscopic models are useful tool for the study of the
 structure of light nuclei, but they can not yield the wave functions in a
 convenient form for future application in nuclear reactions with these nuclei,
 such as above cupture process. Therefore, it is important to develope
 macroscopic models. These models should provide main properties of the
 structure of the nucleus to be investigated.
 \par There are two alternative local potential models, describing the
 alpha-alpha data. The first potential suggested by Ali and Bodmer (AB)
 \cite{ali66} involves a strong repulsive core. The second deep potential
 model was proposed in \cite{buck77} by Buck, Friedrich and Wheatley (BFW),
 which has three forbidden states in the lowest S- and D- waves. Thus, these
 alternative models differ each from other in describing the Pauli repulsion
 part of the alpha-alpha interaction. As a result, the local on-shell
 equivalent potential models give still different wave functions for the
 $^8Be$ ground state: while the BFW potential yields a nodal behavior, the
 AB potential does not describe this microscopically substantiated property.
 \par Macroscopic three-alpha models also are believed to describe well
 the low-lying states of the $^{12}C$ nucleus due-to large binding energy
 of $\alpha$ particle. However, in sharp contrast to the microscopic models,
 the local AB potential gave a very weak binding with $E=-0.60$  \cite{tur97}
for the g.s. energy $E_{exp}=-7.27$ \cite{ajz88} of the $^{12}C$ nucleus as
a srtong off-shell effect of the repulsive core presented in this potential
model. On the other hand, when using the BWF alpha-alpha potential, one has
the problem of elimination of forbidden states from relative motion wave
function of the system. However, the construction of the many body projector
is a very complicated procedure even for the case of the three body system.
Therefore, various approaches to the full many-body projector are used.
\par The approach used in the work \cite{mar82} is based on the strong
truncation of the functional space of relative motion. In spite of the
truncation of the relative motion space, the calculated value overbound
 the ground state binding energy of the $^{12}C$ nucleus by about 5 MeV.
The inclusion of full relative motion space yields a stronger overbinding
by about 15 MeV \cite{wal87} with neglecting the Coulomb forces, which is
estimated by 5-6 MeV. To overcome the three-alpha overbinding problem the
authors of the work \cite{wal87} proposed the Finite Pauli repulsion model
(FPRM) for the $\alpha \alpha$ interaction. However, this model allows
admixtures of the forbidden states into the physical relative motion space.
\par On the other hand, from above mentioned $3\alpha$ macroscopic calculations
with orthogonality conditions, one can not see, how accurate the full 3-body
projector is approximated by chosen orthogonalization method.
Since the antisymmetrization has strong dynamical effects, the $3\alpha$
problem has to be investigated in a model, which allows to examine the
convergence of the orthogonalizing procedure. In other words, the dynamical
effects of the Pauli principle has to be described properly.
\par In the present work we investigate the $3\alpha$-problem by using the BFW
 potential in the framework of the multicluster dynamical model \cite{vor82}
 for light nuclei based on a high accuracy variational method which has
 been used successfully in many structure calculations of various atomic and
 nuclear systems \cite{tur94,tur96,tur98,var95,kam89}.
  For the elimination of forbidden
 states we use the method of orthogonalizing pseudopotentials (OPP) which
 allows to work in the full relative motion space.
 The main feature of our work is the description of
 the convergence of the orthogonalizing procedure for the ground and lowest
 excited states of the $^{12}C$ nucleus.
  When using the OPP method one can examine the convergence of the
 three-body energy as a function of the projecting constant $\lambda$
 $\cite{tur98,baye}$. We check also the the convergense in respect to the
 description of the two-body forbidden states fixed by chosen $\alpha \alpha$
 potential. We show that the convergence of the orthogonalizing procedure
 has different character for the compact shell-model like bound states and
 for the resonance states with a well developed cluster like structure.

\par In Section 2 we describe the model for the investigation of the
 3$\alpha$-system structure. The numerical results are
  presented in Section 3. Discussion and conclusion are given in Section 4
and Section 5, respectively.
  \section { Model  }
 \par For the description of the lowest states of the $^{12}C$ nucleus we
  use the multicluster 3$\alpha$ model with a local $\alpha \alpha$-interaction
potential of Buck, Friedrich and Wheatley \cite{buck77}
\begin{equation}
V(r)=V_0 exp(-\eta r^2),
\end{equation} with $V_0$=-122.6225 MeV, $\eta=0.22$
fm$^{-2}$ \cite{mar82}.  This potential describes well
 the experimental phase shifts of the $\alpha \alpha$-scattering $\delta_L(E)$
with $L=0,2,4$ up to 40 MeV energy. The Coulomb
interaction potential in our calculations is taken in the form
 \begin{equation}
 V_{Coul}(r)= 4e^2erf(br)/r,
 \end{equation}
where b=0.75 fm$^{-1}$,
which corresponds to the $\alpha$ particle charge distribution being the
Gaussian form with a width of $1/b$. We use a value
$ \hbar^2/m_{\alpha}=10.4469$ MeV fm$^2$ in our calculations.
  \par
 The potential (1), with including the Coulomb interaction in the
 form (2) has three nonphysical bound states forbidden by the Pauli
 principle in each $\alpha \alpha$-subsystem, with the energies
 $E(0_1^+)=-72.6249149$ MeV , $E(0_2^+)=-25.6174$ MeV, $E(2^+)=-21.9991037$
 MeV.  The $0_1^+$ state corresponds to the shell configuration $s^8$,
 while the $0_2^+$ and  $2^+$ states correspond to the $s^6p^2$. These
forbidden states in our three-body calculations are eliminated by using the
method of orthogonalising pseudo-potentials (OPP). This method was  developed
in the work \cite{kuk78} and successfully employed for the investigations of
the structure of nuclei with
 $A=6$ and ${\bf 3N}$ system $\cite{tur96,tur98,baye}$.
\par  The variational method on a Gaussian basis used for the calculations of
the spectrum of the ground and excited states of the $^{12}C$ nucleus has been
given in details in Ref.\cite{tur94}.  A high accuracy of the method has been
 demonstrated in a number of works \cite{tur94,tur96,tur98,var95,kam89}.
 A general formalism of the method and analytical expressions of the
 matrix elements of the overlapping integral, the kinetic energy operator,
central, spin-orbital and tensor interaction potentials for the system of three
identical fermions with a spin value 1/2 have been given in Ref.\cite{tur94}.
A corresponding part of these matrix elements can be used
for the variational calculations of the 3$\alpha$-system.  Therefore,
we give here only main formulas for the expression of the wave function
of the 3$\alpha$-system in series of symmetrized Gaussian basis functions.
 \par
The three-body Schr$\ddot o$dinger equation is solved by using a
pseudopotential of the form
\begin{equation}
\widetilde{V}_i(r)=V_i(r)+V_{coul,i}(r) +\sum_f\lambda_f \hat{\Gamma}_i^{(f)},
\end{equation}
where  $\lambda_f$ is the projecting constant,
$\hat{\Gamma}_i^{(f)}$ is the projecting operator to the $f$-wave forbidden
state in the two-body subsystem $(j+k)$, $(i,j,k)=(1,2,3)$, and their cyclic
permutations. We note that the method of OPP for the
elimination of forbidden states uses the first term of the expansion for the
full three-body projector \cite{smir74}
\begin{equation}
\hat{P}=\sum_{i=1}^{3} \hat{P}_i- \sum_{i,j=1}^{3} \hat{P}_i\hat{P}_j+
\sum_{i,j,k=1}^{3} \hat{P}_i\hat{P}_j \hat{P}_k- \cdots,
\end{equation}
where
\begin{equation}
\hat{P}_i=\sum_{f} \hat{\Gamma}_i^{(f)},
\end{equation}
 which corresponds exclusively to the two-cluster Pauli forces.
We neglect three-cluster (triple) Pauli forces when using this approach.
The method of OPP allows us  to obtain the solution of the
Schr$\ddot o$dinger equation at large values of $\lambda_f$ which is
orthogonal to the two-body forbidden states.
  \par Thus, we
use next three-body pseudo-hamiltonian in our 3$\alpha$-cluster variational
calculations:
 \begin{equation} \tilde{H}=H_0
+\tilde{V}_1+\tilde{V}_2+\tilde{V}_3,
\end{equation}
where $H_0$ is the kinetic energy operator of the three $\alpha$- particles.
The wave function of the 3$\alpha$- system is expanded in the series of
symmetrized Gaussian functions \cite{tur94}:
\begin{equation}
\Psi_s^{JM}=\sum_{\gamma j} c_j^{(\lambda ,l)}\varphi_{\gamma j}^s ,
\end{equation}
where $\varphi_{\gamma j}^s=\varphi_{\gamma
j}(1;2,3)+\varphi_{\gamma j}(2;3,1)+ \varphi_{\gamma j}(3;1,2) ,$
\begin{equation} \varphi_{\gamma
j}(k;l,m)=N_jx_k^{\lambda}y_k^lexp(-\alpha_{\lambda j} x_k^2-\beta_{l
j}y_k^2){\cal F}_{\lambda l }^{JM} (\widehat{\vec{x}_k},\widehat{\vec{y}_k})
\end{equation}
Here $(k;l,m)={(1;2,3),(2;3,1),(3;1,2)}$, $\gamma =(\lambda ,l,J,M)=(\gamma_0,J,M);
\vec{x}_k, \vec{y}_k $ are the normalized Jacobi coordinates in the
$k$-set:  $$ \vec{x}_k=\frac{\sqrt{\mu}}{\hbar} (\vec{r}_l-\vec{r}_m)\equiv
\tau^{-1}\vec{r}_{l,m} ; $$ \begin{equation}
\vec{y}_k=\frac{2\sqrt{\mu}}{\sqrt{3}\hbar} (\frac{\vec{r}_l+
\vec{r}_m}{2}-\vec{r}_k)\equiv \tau_1^{-1}\vec{\rho}_k ,
\end{equation}
$N_j$ is a normalizing multiplier. The nonlinear variational parameters
 $\alpha_{\lambda j}, \beta_{l j}$ are chosen as the nodes of the
Chebyshev grid:
$$ \alpha_{\lambda j}=\alpha_0tg(\frac{2j-1}{2N_{\lambda}}\frac{\pi}{2}),
j=1,2,...N_{\lambda}, $$
\begin{equation}
\beta_{l j}=\beta_0tg(\frac{2j-1}{2N_{l}}\frac{\pi}{2}), j=1,2,...N_{l},
\end{equation}
where $\alpha_0$ and $\beta_0$ are scale parameters for each $(\lambda l)$
partial component of the complete wave function. When we use the Chebyshev
grid the basis frequensies  $\alpha_{\lambda j}, \beta_{l j}$ cover larger and
larger intervals around the scale parameters as the numbers $N_{\lambda}$ and
$N_{l}$ increase. This allows us to take into account both short-range and
long-range correlations of particles. The extraordinary flexibility of the
many-particle Gaussian basis makes it possible to describe three- and
four-particle configurations that are formed in the ground state and excited
states of multicluster systems, and which exhibit an extremely high degree of
clustering \cite{tur94}.

The angular part of the Gaussian basis (8) is factorized into the angular
component and the internal wave functions $ \phi(i) $ of the
$\alpha$-particles and has the form:
\begin{equation}
{\cal F}_{\lambda l}^{JM}(\widehat{\vec{x}_k},\widehat{\vec{y}_k})=
\{Y_{\lambda}(\widehat{\vec{x}_k}) \bigotimes Y_l(\widehat{\vec{y}_k})\}_{JM}
 \phi(1) \phi(2) \phi(3)
\end{equation}
Here the orbital momenta $\lambda$ and $l$ are conjugate to the Jacobi
coordinates $\vec{x}_k$ and $\vec{y}_k$, respectively.
 \par  We note that the use of the symmetrized Gaussian basis allows us to
take into account contributions actually of all partial waves to the binding
energy of the system due to overlapping of
various three-body channels with the same total orbital momentum ${\bf L}$ (
and total spin ${\bf S}$  in the case of the ${\bf 3N}$-system). On the other hand,
the solution of the Schr$\ddot o$dinger equation should be symmetric over the
permutation of any two ${\bf \alpha} $ particles. This requirement is
satisfied automatically when using a symmetrized basis for any number of basis
functions, whereas in the case of nonsymmetric basis this requirement is
fullfilled approximately and depends on the precision of the expansion.
\par One of the main advantages of the variational method on a Gaussian basis
is the possibility of fully analytical calculation of moments of interaction
 potentials of some special forms. For example, to calculate the integral
 \begin{equation}
 G(\alpha,\beta,n)= \int_0^{\infty} x^n exp(-\alpha x^2-\beta x) dx ,
\alpha >0, n \geq 0,
 \end{equation}
 which arises when Yukawa type potentials are used, we suggest more
 convenient and effective recurent formulas (compare with Ref.\cite{var95}):
 \begin{eqnarray}
  G(\alpha,\beta,0)= \frac{\sqrt {\pi}}{2 \sqrt{\alpha}}
 exp(\frac{\beta ^2}{4 \alpha}) erfc(\frac{\beta}{2 \sqrt{\alpha}}) \\
G(\alpha,\beta,1)= \frac{1}{2 \alpha} -\frac{\beta}{2\alpha} G(\alpha,\beta,0)
 \\ G(\alpha,\beta,n+1)= \frac{n}{2 \alpha}G(\alpha, \beta, n-1)
 -\frac{\beta}{2\alpha} G(\alpha,\beta,n), n >0
 \end{eqnarray}
 And for the calculation of moments of the nonpoint Coulomb interaction
 potential (2)
 \begin{equation}
 G_{erf}(\alpha,\beta,n)= \int_0^{\infty} x^n exp(-\alpha x^2) erf(\beta x)dx
 , \alpha >0, n \geq 0,
 \end{equation}
 we obtained also next recurent formulas:
 \begin{eqnarray}
  G_{erf}(\alpha,\beta,1)= \frac{\beta}{2 \alpha \sqrt{\alpha+\beta^2 }} \\
G_{erf}(\alpha,\beta,n)= \frac{n-1}{2 \alpha}G_{erf}(\alpha, \beta,
 n-2) +\frac{\beta}{2\alpha \pi} \frac {\Gamma(n/2)}{(\alpha
 +\beta ^2)^{n/2}}
  \end{eqnarray}
 These formulas allow us to essentially economize computer time when we
tabulate the values of $ G(\alpha,\beta,n)$ and $G_{erf}(\alpha,\beta,n)$
 for the fixed values of $\alpha$ and $\beta$.
We note also that the zero-th moment $G_{erf}(\alpha,\beta,0)$ can be
 estimated numerically. However, in our calculations we need moments with only
odd values of $n$.
\section { Numerical results }
\par First we note that a very large symmetrized
Gaussian basis is used in our calculations. For the calculations of the energy
levels with  $(J^{\pi},T)=(0^+,0)$ we used a basis containing 80 functions in
 the each of the three-body channels $(\lambda,l)=\{ (0,0); (2,2); (4,4)\}$.
We calculated the levels with $(J^{\pi},T)=(2^+,0)$ on the basis containing
 437 Gaussians chosen in main three-body channels $(\lambda,l)=\{ (0,2); (2,0);
 (2,2); (2,4); (4,2); (4,4) \}$.  The results indicate that a further
extension of the basis does not have a remarkable influence on the accuracy of
the expansion.
 \par The projector on the $\bf f$-wave forbidden state in each
two-body subsystem has the form:  $$ \hat{\Gamma}_i^{(f)}=\frac{1}{2f+1}
\sum_{m_f} \mid \varphi_{f m_f}(\vec{x}_i)> < \varphi_{f m_f}(\vec{x'}_i) \mid
\delta(\vec{y}_i-\vec{y'}_i) , $$ with the forbidden state function $$
\varphi_{f m_f}(\vec{x}_i) =x_i^f \sum_m N_m^{(f)}b_m^{(f)}exp(-\frac{r_i^2}{2r_{0m}^{(f)2}})
Y_{fm_f}(\hat{\vec{x}_i}).  $$
Here $r_0$ is the "projector
radius", and $N_m^{(f)}$ is the normalizing multiplier:  $$ N_m^{(f)}=2^{f+7/4}
\frac{\alpha_m^{(2f+3)/4}}{\pi ^{1/4}[(2\lambda+1)!!]^{1/2}}, \qquad
\alpha_m=\tau^2/(2 r_{0m}^2).  $$
 \par In order to check the behaviour of the 3$\alpha$-system energy when
improving the accuracy of the expansion of the two-body $\alpha \alpha$ forbidden
states wave functions in a series of Gaussian functions we choose several
sets of the Gaussian approximations. Corresponding approximate values of the
 deep forbidden state energies of the $\alpha \alpha$-system are shown
in the Table 1. In Set 1 the forbidden state wave function of the
 $0_1^+$-level of the $\alpha \alpha$-system is approximated via  N=1
Gaussian, and wave functions of the forbidden  $0_2^+$- and $2^+$-levels
are approximated via N=2 Gaussians, et al.  The values of the forbidden
two-body states energies are given in corresponding squares. By comrarison of
these numbers with the corresponding exact values of the two-body
$0_1^+, 0_2^+, 2^+$-forbidden states energies one can conclude about the
quality of the approximation for the given number N.  We note that the wave
function of the $\alpha\alpha$ forbidden $0_2^+$-state
contains a node, therefore it can not be expressed by a single Gaussian.
\par The spectrum of the energy levels of the $^{12}C$ nucleus with  $(J^{\pi},
T)=(0^+,0)$ and $(2^+,0)$ for the several variants
of the Gaussian expansion for the forbidden state wave functions are
 shown in Tables 2 and 3. For the sake of convenience we use
 an identical value of the projecting constant $\lambda_f$ for all forbidden
 states. In all three tables the symbol "$<P>$" denotes the total admixture of
 the forbidden states to the energy of the 3$\alpha$-system:
 $$ <P> = < \Psi_s \mid
 \lambda(\hat{P_1}+\hat{P_2}+\hat{P_3}) \mid \Psi_s >. $$
\par In Table 4 we give the experimental values of the lowest energy levels
 of the $^{12}C$ nucleus and our theoretical estimation comparing with
 results from the literature. The results of the Ref.$\cite{mar82}$ were
obtained by using the hyperspherical
garmonics method and the BFW $\alpha\alpha$-potential. Microscopical
calculations in the Ref.$\cite{cso97}$ were performed with an effective
MN nucleon-nucleon potential.
\par
From Tables 2 and 3 one can see that we have a reasonable saturation of the
energy levels of the $^{12}C$ nucleus when increasing the projecting constant
$\lambda$ to the infinity for a given set of approximation of the two-body
forbidden states.
  However, beginning from the value of the projecting constant
$\lambda \geq 10^4$ MeV the ground and first excited $2_1^+$ states energies
of the $^{12}C$ nucleus begin to display a high sensitivity to the description
 of the two-body forbidden states wave functions.
When improving a quality of the approximation of the $\alpha \alpha$-forbidden
states in these cases we go to the weakly bound system .
\par For the other excited states ($0_2^+, 0_3^+,2_2^+$)of the $^{12}C$
nucleus we have a quite reasonable convergence of the energy with respect to
the description of the $\alpha \alpha$-forbidden states.
\par We note also that admixture of the forbidden states in the ground and
first excited $2_1^+$ states of the $^{12}C$ nucleus decreases
slowly as increasing the projecting constant $\lambda$ to the infinity.
 In the case of the $3N$ system at $\lambda=10^5$ MeV
the value of the forbidden states admixture in the $^3H$ ground state energy
was approximately $10^{-3}$ MeV when using the Moscow $NN$ potential with
forbidden states $\cite{tur96,tur98}$.

\section { Discussion}
\par  First of all we discuss the reason of the strong dependence of the
 energies of the ground and excited $2^+_1$ states of the $^{12}C$ nucleus
 on the expansion accuracy of the two-body forbidden states wave functions.
  We know from the experiment that these states are bound strongly and
 have consequently a well developed compact shell-model structure \cite{ajz88}.
Thus, in the states with $(J^{\pi},T)=(0^+_1,0)$ and $(J^{\pi},T)=(2^+_1,0)$
three $\alpha$-clusters with a large probability overlaps simultaneously each
with other. In other words,  three-cluster Pauli forces should play an
important role in the description of these states due to the requirement of the
Pauli principle. However, the method of OPP used here for the elimination of
forbidden states in the full three-body system takes into account really
two-body Pauli effects.
\par It is known that the two-body projectors $\hat{P_i}$ and  $\hat{P_j}$
do not commute each with other and are not mutually orthogonal \cite{smir74}.
We suggest that the overlapping of these two-body projectors is very strong
for the three-alpha system, and hence the energies of the compact ground
 and excited $2^+_1$ states of the $^{12}C$ nucleus display a high
 sensitivity to the description of the forbidden states beginning from
$\lambda \approx 10^4$ MeV where the next terms in the expansion (11) begin
to play a role. And these neglected terms present a mathematical expression of
the Pauli triple  forces.
\par We note that the authors of the Ref.\cite{oryu94} also came to
the conclusion that three-body Pauli forces should be taken into account
when investigating the structure of the 3$\alpha$-system. They made this
conclusion on the basis of the calculations of the $^{12}C$ nucleus ground
state  energy in the framework of the Alt-Grassberg-Sandhas equation method by
using $\alpha \alpha$-potentials obtained by the RGM and modified by the OCM.
 \par On the other hand, as we noted in the previous section, the energies of
the excited levels with $(J^{\pi},T)=(0^+_2,0)$ and $(J^{\pi},T)=(0^+_3,0)$
change slowly from Set 1 ($N \leq 2$) to Set 4 ($ N=7$). Moreover, at $\lambda
\approx 10^7 - 10^8$ MeV one can see a good saturation of the energy of these
levels with respect to values of the projecting constant $\lambda$. Thus, for
the energies of these levels we can write (see Table 2) :
$E_{theor}(0^+_2)=1.551$ MeV and $E_{theor}(0^+_2)=4.055$ MeV. It is important
to note that the difference of these values $\Delta_{theor}= 2.50$ MeV is in a
good agreement with it's experimental value $\Delta_{exp}=2.65$ MeV.
\cite{ajz88}. And this agreement can be seen for all sets 1-4 at large
values $\lambda \approx 10^7 - 10^8$ MeV.  Thus, the excited levels
 of the $^{12}C$ nucleus with $(J^{\pi},T)=(0^+_2,0)$ and $(J^{\pi},T)=(0^+_3,0)$
 can be described qualitatively in the framework of the method of OPP in
 contradistinction to the ground and first $2^+_1$ excited states.
\section { Conclusions}
\par The energy spectrum of the $^{12}C$ nucleus with $(J^{\pi},T)=(0^+,0)$
and $(J^{\pi},T)=(2^+,0)$ was calculated in the framework of the multicluster
dynamical model using the $\alpha \alpha$-potential of Buck, Friedrich and
 Wheatley with forbidden states in the $S$ and $D$ waves. For the
  elimination of forbidden states in the full three-body system the method of
  orthogonalising pseudo-potentials (OPP) has been used.
  \par It was shown that the energies of the compact $(0^+_1,0)$ and
  $(2^+_1,0)$ states of the $^{12}C$ nucleus with a shell-model like
  structure display a very high sensitivity
  to the description of the two-body forbidden states. Therefore, for the
  energies of these strongly bound states, the chosen model does not give
  convergent results. We suggest that this situation is due-to neglecting
  of three-body Pauli forces, which should play an important role in the
  description of the structure of these strongly bound states.
   \par We obtained also that in our model the resonance states  $(0^+_2,0)$
  and $(0^+_3,0)$ with a well developed cluster-like structure  can be described
  quite well. In contrast to the ground and first excited states, the
  orthogonalization procedure yields convergent results. However, the energy
  position of the astrophysically significant $0_2^+$ resonance
  $E_{exp}(0^+_2)=0.3796$ MeV is overestimated by about 1.2 MeV.

 \begin{center}
{\bf ACKNOWLEDGEMENTS}
  \par Author is thankful to Prof. D.Baye, Prof. A.Mukhamedzhanov,
Prof. A.Csoto and Dr.G.Kim for useful discussions, and Dr. G.Ryzhikh and
Dr. K.Varga for comparison and discussion of presented results. This work
was supported in part by DAAD.
\end{center}

 \newpage
\begin{table}
\caption { Sets of Gaussian approximations for the forbidden states wave
functions of the $\alpha \alpha$ system and corresponding energy values in MeV}
\begin{tabular}{|c|c|c|}       \hline
$0^+_1 $              &$0^+_2$                &  $2^+$  \\    \hline
exact $E=-72.624915$&exact $E=-25.6174$ & exact $E=-21.999104$  \\  \hline
$ N=1: E=-72.5445$   &$ N=2: E=-25.106$   & $ N=2: E=-21.676 $  \\  \hline
$ N=3: E=-72.6126$   &$ N=3: E=-25.5558$  & $ N=3: E=-21.8837 $  \\  \hline
$ N=4: E=-72.6233$   &$ N=4: E=-25.6111$  & $ N=4: E=-21.9576 $  \\  \hline
$ N=7: E=-72.624905$   &$ N=7: E=-25.6173$  & $ N=7: E=-21.999098 $  \\ \hline
\end{tabular}
\end{table}
\begin{table}
\caption { The energy spectrum of the  $^{12}C$ nucleus with
 $(J^{\pi},T)=(0^+,0)$ in MeV for the several sets of Gaussian approximations
 of the two-cluster forbidden states at several values of the
 projecting constant}
 \begin{tabular}{|c|c|c|c|c|c|c|c|c|c|}       \hline
&$\lambda$ (MeV)       &10      &$10^2$ &$10^3$ &$10^4$ &$10^5$ &$10^6$ &$10^7$ &$10^8$ \\   \hline
$N\leq 2$& $E_1$ & -210.80&-45.194&-20.665&-18.999&-14.882&-12.85&-12.56 &-12.44 \\
         & $E_2$ & -150.72&-20.037&-0.677 &+0.274 &1.210  &1.284 &1.39   &1.42 \\
         & $E_3$ & -109.37&-15.878&+1.239 &1.306  &+2.98  &3.446 &3.99   &4.01 \\
         & $<P>$ & 29.83  &30.36  & 0.847 &1.133  &1.73   &0.300 &6.E-2  &5.E-2 \\  \hline
         $N=3  $  & $E_1$ & -210.71&-43.298&-19.776&-16.492&-6.387&-3.666&-3.35 &-3.311 \\
         & $E_2$ & -150.05&-17.183&-0.411 &-0.021 &1.238  &1.331 &1.42   &1.463 \\
         & $E_3$ & -109.53&-15.652&+1.344 &1.380  &+2.553  &3.361 &4.07   &4.046 \\
         & $<P>$  & 29.93  &28.78  & 1.017 &2.837  &3.100   &0.322 &1.E-2  &1.E-2 \\  \hline
$N=4  $  & $E_1$ & -210.72  &-44.149  &-20.15  &-16.227  &-.463    &0.051  &0.406   &0.468 \\
         & $E_2$ & -150.28  &-17.423  &-0.512   &-0.417   &1.357   &1.424   &1.522  &1.571 \\
         & $E_3$ & -109.37  &-15.430  &+1.336   &1.355    &+2.979  &3.319   &4.059  &4.087 \\
         & $<P>$  & 29.92    &28.76    & 1.120   &3.647    &0.397   &0.242   &6.4E-2 &8.7E-3 \\  \hline
$N=7  $  & $E_1$ & -210.69  &-44.207  &-20.15  &-16.106  &-.830    &-0.435  &-0.307 &-0.283 \\
         & $E_2$ & -150.32  &-17.585  &-0.531   &-0.422   &1.353   &1.407   &1.513  &1.551 \\
         & $E_3$ & -109.32  &-15.310  &+1.334   &1.353    &+3.019  &3.316   &4.038  &4.055 \\
         & $<P>$  & 29.95    &28.92    & 1.130   &3.777    &0.721   &8.76E-2
         &2.3E-2 &4.7E-3 \\  \hline
\end{tabular}
\end{table}

\newpage
\begin{table}
\caption { The same for the levels with $(J^{\pi},T)=(2^+,0)$ }
\begin{tabular}{|c|c|c|c|c|c|c|c|}       \hline
&$\lambda$ (MeV) &$10^3$   &$10^4$  &$10^5$  &$10^6$  &$10^7$  &$10^8$ \\   \hline
$N\leq 2$& $E_1$ &-17.19   &-16.15  &-14.55  &-12.103 &-11.176 &-11.04   \\
         & $E_2$ & 0.944   &0.95    &0.952   &0.965   &1.022   &1.079   \\
         & $<P>$  & 0.923   &0.95    &1.024   &0.807   &0.136   &2.E-2   \\   \hline
$N= 3$   & $E_1$ &-17.213   &-15.665  &-11.142  &-5.953 &-3.950 &-3.80   \\
         & $E_2$ & 1.029    &1.032    &1.034   &1.047   &1.141   &1.19   \\
         & $<P>$  & 1.024    &0.918    &2.842   &1.495   &5.E-2   &8.5E-3   \\   \hline
$N= 4$   & $E_1$ &-17.433   &-15.725  &-8.578  &1.042  &1.082   &1.157   \\
         & $E_2$ & 1.029    &1.032    &1.034    &1.694   &2.656   &3.038   \\
         & $<P>$ & 1.075     &1.095    &5.85      &9.1E-3   &2.9E-2&2.8E-2   \\   \hline
$N= 7$   & $E_1$ &-17.361   &-15.649  &-8.243   &1.042  &1.086   &1.162   \\
         & $E_2$ & 1.030    &1.032    &1.034    &1.475   &2.524   &2.643   \\
         & $<P>$ & 1.062     &1.127    &5.974    &1.2E-2   &3.1E-2 &2.7E-2
\\ \hline
\end{tabular}
\end{table}
\begin{table}
\caption { Experimental results and theoretical estimations of energies of the
  ground and lowest excited states of the $^{12}C$ nucleus in MeV}
\begin{tabular}{|c|c|c|c|c|}       \hline
     & Exp. & BFW $\alpha\alpha$-potential& MN effective $NN$-potential &
 BFW $\alpha\alpha$-potential \\
     &$\cite{ajz88}$ & $\cite{mar82}$& $\cite{cso97}$ & with OPP \\ \hline
$0_1^+$&-7.275 & -12.5 &-10.43  & not converged  \\
$0_2^+$&0.3796 & -2.7  &0.64    & 1.55  \\
$0_3^+$&3.0    &  --    & 5.43   & 4.055   \\
$2_1^+$&-2.836 & -10.0 &-7.63   & not converged  \\
$2_2^+$&3.89 & -3.7 &6.39 &2.64  \\ \hline
\end{tabular}
\end{table}
\end{document}